\documentclass[prl, twocolumn, superscriptaddress]{revtex4-2}
\usepackage[version=3]{mhchem}
\usepackage[normalem]{ulem}
\usepackage{graphicx} 
\usepackage{lastpage}
\usepackage{fnpos}
\usepackage[english]{babel}
\addto{\captionsenglish}{%
  \renewcommand{\refname}{Notes and references}
}
\usepackage[T1]{fontenc}
\usepackage[usenames,dvipsnames]{xcolor}
\usepackage{hyperref}

\usepackage{epstopdf}

\usepackage{booktabs}

\definecolor{cream}{RGB}{222,217,201}
\definecolor{darkred}{rgb}{0.9098 , 0.2824, 0.3333}
\newcommand{\NB}[1]{\textcolor{black}{#1}}
\newcommand{\ML}[1]{\textcolor{black}{#1}}

\usepackage[utf8]{inputenc}
\usepackage{siunitx}
\usepackage{tikz}
\usetikzlibrary{arrows,shapes,positioning}
\usetikzlibrary{decorations.markings,decorations.pathmorphing,
decorations.pathreplacing}
\usetikzlibrary{calc,patterns,shapes.geometric, matrix}

\DeclareSIPrefix\micro{\text{\textmu}}{-3}

\usepackage{pgfplots}
\pgfplotsset{compat=1.5}

\usetikzlibrary{external}
\graphicspath{ {tikz/}, {head_foot/} }

\begin{document}

\title{Capillary and priming pressures control the penetration of yield-stress fluids through non-wetting 2D meshes}

\author{Manon Bourgade}
\affiliation{Universite Claude Bernard Lyon 1, CNRS, Institut Lumière Matière, UMR5306, F-69100, Villeurbanne, France}
\affiliation{Saint-Gobain Recherche Paris, 39 Quai Lucien Lefranc, 93300 Aubervilliers, France.}

\author{Nicolas Bain}
\author{Loïc Vanel}
\author{Mathieu Leocmach}
\author{Catherine Barentin}
\affiliation{Universite Claude Bernard Lyon 1, CNRS, Institut Lumière Matière, UMR5306, F-69100, Villeurbanne, France}
\email{catherine.barentin@univ-lyon1.fr}

\begin{abstract}
Forcing hydrophilic fluids through hydrophobic porous solids is a recurrent industrial challenge.
If the penetrating fluid is Newtonian, the imposed pressure {has to overcome the capillary pressure at the fluid-air interface in a pore}. The presence of a yield-stress, however, makes the pressure transfer {and the penetration} significantly more complex.
In this study, we experimentally investigate the forced penetration of a {water based} yield-stress fluid through a regular hydrophobic mesh under quasi-static conditions, combining quantitative pressure measurements and direct visualisation of the penetration process.
We reveal that the penetration is controlled by a competition between the yield-stress and two distinct pressures.
The capillary pressure, that dictates the threshold at which the yield-stress fluid penetrates the hydrophobic mesh, and a priming pressure, that controls how the fluid advances through it.
The latter corresponds to a pressure drop ensuing a local capillary instability, never reported before.
Our findings shine a new light on forced imbibition processes, with direct implications on their fundamental understanding and practical engineering.
\end{abstract}

\maketitle

\section{Introduction}

{We all intuitively know how to force a fluid into a porous media.
Take a sponge for instance, and use it to wipe some fluid off a surface.
Whether the sponge is dry or not, with small or large pores, and whether that fluid is water, oil, or some thick sauce, we apply a different pressure to absorb it.
In other words, the pore geometry, the fluid rheology, and its capillary affinity with the porous media all matter.
{When the medium is fully saturated with liquid, Darcy law \NB{describes} the viscous resistance against a forced flow.}
\ML{In this case, the flow rate is proportional to the applied pressure difference.}
When \ML{the porous medium} is not saturated, capillarity kicks in.
\NB{If the fluid wets the medium, it is spontaneously absorbed by it \cite{deng2024modeling}.}
\NB{If it does not wet the medium},
\ML{
the capillary pressure at the fluid-air interface acts as a \emph{threshold pressure} that must be overcome to observe any flow~\cite{haines1930studies,armstrong2013interfacial, moebius2012interfacial,sun2019haines}.}
{Such capillary constraints lead to a penetration behaviour much more intricate than the steady-state Darcy flow~\cite{haines1930studies, armstrong2013interfacial, moebius2012interfacial,sun2019haines}}.
{When the fluid has a non-Newtonian rheology, the penetration behaviour is even more complex and remains scarcely investigated.}}

{Out of the kitchen, however, the forced imbibition of non-Newtonian fluids into porous media is crucial for many practical applications, including filtration, textile processing and washing, or civil engineering.
In particular, {water-based yield-stress fluids \cite{bonn2017yield} {such as pastes, polymeric or colloidal gels} represent many everyday fluids and are ubiquitous in industrial context.
The control of their penetration inside \emph{hydrophobic} porous media, e.g. filters, fabric, skin or construction materials is a recurrent issue.}
}

\ML{In wetting situations}, the spontaneous imbibition of porous media by Newtonian fluids has been an active topic for a long time \cite{Peek}. Capillary absorption or water transport by textiles has been widely studied{, and in particular the role of liquid saturation\cite{Ghali}, porosity scales~\cite{Pezron}, contact angle~\cite{Ezzatneshan} and geometric details~\cite{Duprat}.}
{When it comes to \emph{yield-stress} fluids, experimental and numerical investigations of the forced flow into homogeneously filled porous and fibrous media showed a behavior consistent with a modified Darcy law, {where the yield stress induces another threshold pressure below which the fluid does not flow} \cite{Chevalier, Shahsavari, bauer2019experimental, chaparian2024yielding, chaparian2025comprehensive}.} 
\ML{This threshold pressure induced by yield stress, different in nature from the capillarity induced one, was also observed in falling drop experiments \cite{blackwell2016impacts}.}

\ML{In non wetting situations, the study of forced penetration has been limited to Newtonian fluids. It has been investigated both at a single pore level~\cite{Zong, Lee_2009}, and in fiber layers \cite{jamali2019penetration}, either by falling drop experiments \cite{Delbos, Zhang, Zong, Bae, Ryu, abouei2022spectacular} or by static pressures \cite{jamali2019penetration,Zong}.
All evidenced the existence of a capillary threshold pressure, influenced by geometry, below which penetration does not occur.
The forced penetration of yield-stress fluids, and how yield stress combines to the capillary threshold pressure, therefore remain unexplored.}

{In this study,
{we address this challenge by experimentally investigating the penetration behaviour of a water-based yield-stress fluid into a hydrophobic fibrous mesh with a quasi-static forced imbibition setup.}
{After presenting our experimental setup, we first report measurements of the threshold penetration pressure required for a yield-stress fluid to pass through a hydrophobic mesh.}
{We show that in the range of parameter we explore the threshold penetration pressure is dictated by the capillary pressure and has little dependence on the yield stress.
We then focus on the local penetration phenomenology.
In stark contrast, we show through detailed observation and modeling that the yield stress has a profound impact on the microscopic instabilities that govern the penetration path.}

\begin{figure}[t]
\centering\includegraphics[width=\columnwidth]{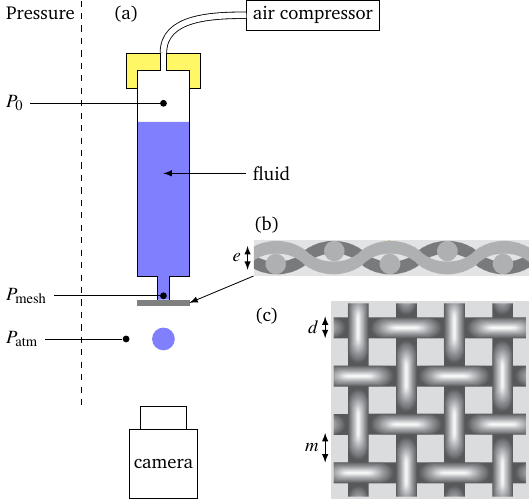}
  \caption{(a) Quasi-static experimental setup. (b) and (c) Side-view and top-view diagrams, respectively, of the woven polyamide meshes (diagrams provided by the manufacturer). Definition of key dimensions: pore size $m$, fiber diameter $d$, and mesh thickness $e$.}
  \label{fig:setup}
\end{figure}

\section{{Experimental setup}}

\subsection{{Quasi-static forced imbibition}}

\noindent
We illustrate {our} quasi-static forced imbibition setup in Fig.~\ref{fig:setup}. 
We fill a syringe with a model yield-stress fluid. 
At one end of the syringe, we glue a hydrophobic mesh (see Supplementary Section 1 for details).
At the other end, we connect an air compressor that applies a controlled pressure $P_0$.
{This leads to a pressure at the mesh interface $P_{\mathrm{mesh}}= P_0 + \Delta P_h - \Delta P_y$, where $\Delta P_h$ is the hydrostatic pressure and $\Delta P_y$ the contribution of the yield-stress fluid in the syringe (see Supplementary Section 2 for details).}
In each measurement, we increase the applied pressure until the fluid flows through the mesh \ML{(see Supplementary Section 3 for detailed experimental steps)}.
\\

We used an \textsc{MFCS} (Microfluidic Flow Control System) air compressor from \textsc{Fluigent}, that can impose a pressure $P_0 =  P_{\mathrm{atm}} + \Delta P_\mathrm{comp}$, where $\Delta P_\mathrm{comp}$ ranges from 0 to \SI{6900}{\pascal} with a \SI{2.5}{\%} precision.
We manually increase this pressure using the software provided by the manufacturer, with increments of \SI{2}{\pascal} at each step, at a slow rate of about \SI{10}{\pascal\per\s} to keep  {the flow} quasi-static.

\begin{table}[t]
    \centering
    \begin{tabular}{@{}crrrr@{}}
                \toprule
			Mesh \# & $m$ (\si{\micro\metre}) & $d$ (\si{\micro\metre}) & $e$ (\si{\micro\metre}) & $m_\mathrm{real}$ (\si{\micro\metre}) \\\midrule
                1& 64 & 33 & 50 & $71\pm2.5$\\
                2& 85 & 24 & 40 & $84\pm1.5$\\
                3& 105 & 40 & 63 & $104\pm2.1$ \\
                4& 125 & 62 & 100 & $118\pm2.3$ \\
                5& 190 & 62 & 100 & $177\pm1.5$ \\
                \bottomrule 
            \end{tabular}
    \caption{Hydrophobic mesh geometric properties. The pore size $m$, fiber diameter $d$ and mesh thickness $e$ are defined in Fig.\ref{fig:setup}, and their values provided by the manufacturer. The true pore size $m_{real}$ corresponds to the value we measured with the macroscope.}
    \label{tab:sefar_mesh}
\end{table}
\subsection{Model yield stress fluids and 2D meshes}

For the model yield-stress fluid\cite{ovarlez2013}, we chose an aqueous suspension of Carbopol polymer microgels (Ultrez 10 powder) from Lubrizol \cite{jaworski} (see Supplementary Section 4 for preparation protocol). 
For polymer concentrations above the jamming threshold $c^{*}\approx 0.09$\% w/w, the suspension exhibits a yield stress resulting from the jamming of the microgels, whereas below $c^*$ the suspension has no yield stress.
To characterize the mechanical properties of the suspensions, we measure the flow curve using a rheometer (Anton Paar Physica MCR 302) with a parallel-plate geometry, and determine the yield stress $\sigma_y$ by fitting the data to a Herschel-Bulkley model (see Supplementary Section 5).
For polymer concentrations ranging from 0.1\% w/w to 1.2\% w/w, the yield stress $\sigma_y$ varies from \SI{4.4}{\pascal} to \SI{86}{\pascal}$\pm 10$\%.

As for the hydrophobic meshes, we selected single layer meshes of woven polyamide fibers (SEFAR), with well-controlled and uniform pore size $m$, fiber diameter $d$, and mesh thickness $e$ (Fig.~\ref{fig:setup}b-c).  
The manufacturer-provided dimensions are given in Table~\ref{tab:sefar_mesh}, and the corresponding references in {Supplementary Table~\ref{tab:sefar_mesh_ref}}.

To verify the meshes hydrophobicity,  we placed water drops on their surface and measured macroscopic contact angles using a side camera and the \texttt{Dropsnake} module in ImageJ~\cite{dropsnake}. 
Measurements from four different drops give a superhydrophobic average contact angle of $\ang{130}$.
In addition, we performed contact angle measurements on individual fibers, for which we estimated a contact angle ranging between \ang{90} and \ang{100} (see Supplementary Section 6 for details).

\begin{figure*}
    \includegraphics[width=\textwidth]{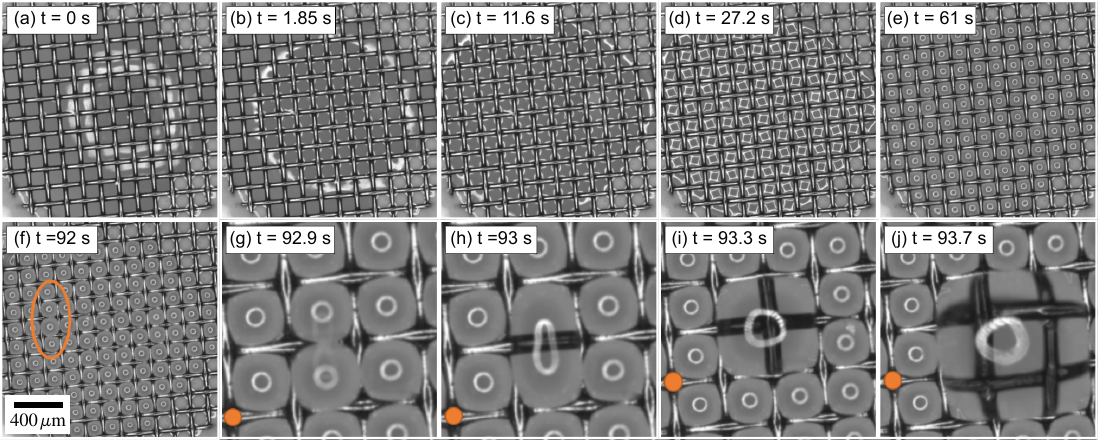}
    
    \caption{Snapshots from an experiment {with mesh 3 and a fluid with a yield stress of \SI{66}{\pascal}}, showing the progression of the fluid-air interface.
    {The camera focus is on the hydrophobic mesh, for which we can observe some side pores filled with glue.}
    {(a) The fluid approaches the mesh. The ring light forms a central circular reflection on the fluid front.}
    (b) The fluid contacts the mesh at the center. The pores here show a small white reflection at the edges, indicating initial contact. At this step, the compressor pressure is $\Delta P_{\mathrm{comp}} = \Delta P_{\mathrm{contact}}$, the initial flowing pressure required to bring the yield stress fluid to the tip of the syringe.
    (c) The fluid contacts the entire mesh.
    (d) The fluid starts to advance into the pores, as the light reflections shift to square shapes in the pore centers.
    (e) The fluid lightly protrudes through the pores, forming circular reflection patterns.
    (f) As we increase the applied pressure, drops expand through the pores, nearing neighboring drops. The circled area indicates where the first breakthrough will occur.
    (g) Neighbouring drops touch and begin coalescing, accelerating the dynamics.
    (h) to (j) Successive coalescences quickly connect multiple pores, forming clusters as the fluid penetrates.
    }
    \label{fig:grille}
\end{figure*}

\subsection{{Penetration visualisation}}

We visualise the fluid-air interface with a ZEISS Axio Zoom V16 microscope placed underneath the quasi-static compression setup (Fig.~\ref{fig:setup}a).
This system, capable of magnifications up to 112x, will be referred to as a macroscope hereafter.
We used it to quantify the true pore size of the hydrophobic meshes $m_{real}$ (Table~\ref{tab:sefar_mesh}), and to record videos of the fluid-air interface throughout the experiments.
We illuminate the fluid-air interface with a ring light, and conveniently exploit its reflection pattern to follow the progression of the fluid through the pores (Fig. \ref{fig:grille}).
{A typical experiment goes as follows.}
{As we increase the applied pressure, the fluid approaches, contacts, and progressively flows through the mesh (Fig.~\ref{fig:grille}a-d), until it protrudes from it (Fig.~\ref{fig:grille}e-f).
Following the first drop coalescence (Fig.~\ref{fig:grille}g), we maintain a constant pressure. Subsequent fusions occur rapidly — within less than \SI{1}{\s} (Fig.~\ref{fig:grille}g-j) — resulting in a connected cluster (Fig.~\ref{fig:grille}j), and eventually passage of the fluid through the mesh.}

Informed by this phenomenology, we define the penetration pressure $\Delta P_{\mathrm{pen}}$ as the difference between the pressure applied by the compressor at the moment of the first coalescence event (Fig.~\ref{fig:grille}g) and the one at which the fluid is brought in contact with the mesh (Fig.~\ref{fig:grille}b).
\NB{Experimentally, we measure it as $\Delta P_{\mathrm{pen}} = \Delta P_{\mathrm{coalescence}} - \Delta P_{\mathrm{contact}}$.}

\section{Results and discussion}

\subsection{The penetration pressure is set by capillarity}

\subsubsection{Penetration pressure for a Newtonian fluid}

{We first consider the simpler case of a Newtonian fluid, in the absence of yield stress.
Once the fluid contacts the mesh, the local curvature of the liquid-air interface in each pore gives rise to a pressure jump, the Laplace pressure $\Delta P_{L}$.
For a square pore formed by cylindrical fibers \cite{jamali2019penetration},
}
\begin{equation}
    \Delta P_{L}(\alpha) =  -4 \, \frac{\Gamma\sin (\alpha +\theta)}{m + \ell(1-\sin \alpha )},
    \label{eq:Jamali}
\end{equation}
{where $\alpha$ is the immersion angle along the fiber, $\Gamma$ the surface tension of the fluid-air interface,  $\theta$ the fluid-fiber contact angle,  $m$ the pore size, and $\ell$ the pore depth (Fig. \ref{fig:jamali_1} inset)}.

{The Laplace pressure $\Delta P_{L}(\alpha)$ defined 
in Eq.~\eqref{eq:Jamali} has a non-monotonic behavior with the immersion angle $\alpha$ (Fig.~\ref{fig:jamali_1}). 
Advancing the meniscus through the pore first increases the Laplace pressure, until it reaches a maximum at a critical immersion angle $\alpha_c$.
Then, the meniscus can progress spontaneously through the pore, without any further pressure increase. 
We therefore expect the pressure at which a Newtonian fluid penetrates the mesh $\Delta P_{\mathrm{pen}}$ to equate the maximal Laplace pressure $\Delta P_{L,\mathrm{max}} = \Delta P_L(\alpha_c)$.}

\begin{figure}
   \centering
    \includegraphics[width=\columnwidth]{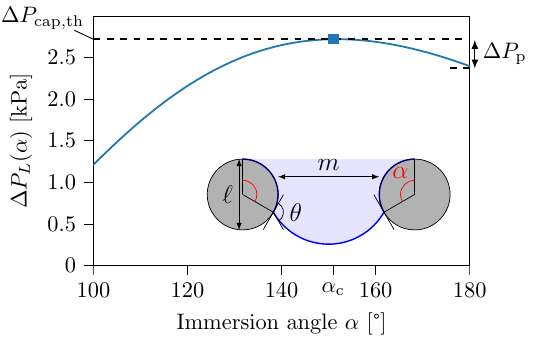}
   
%
%
    \caption{
    Main. Laplace pressure as a function of the immersion angle $\alpha$, for $\theta=90^{\circ}$, $m=\SI{71}{\micro\metre}$ and $\ell=\SI{33}{\micro\metre}$. The maximum pressure $\Delta P_{L,\mathrm{max}}$ is reached at the critical immersion angle $\alpha_c$. The priming pressure $\Delta P_p$ is defined as $\Delta P_p= \Delta P_{L,\mathrm{max}} - \Delta P_L(180^{\circ})$.
    Inset. Definition of the parameters: $\alpha$ is the immersion angle, $\theta$ is the contact angle on a single fiber, $m$ the pore size, $\ell$ the pore depth. 
    }
    \label{fig:jamali_1}
\end{figure}

\begin{figure}
    \centering
        
    \includegraphics[trim = 0.1cm 0 0 0,clip,width=\columnwidth]{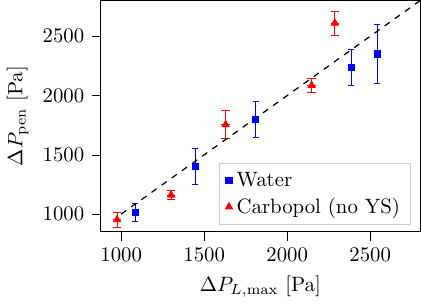}
    
    \caption{Experimentally measured penetration pressure $\Delta P_{\mathrm{pen}}$ as a function of theoretical maximum Laplace pressure $\Delta P_{L,\mathrm{max}}$, obtained with water (blue squares) and dilute Carbopol suspension without yield stress (red triangles).
    {The symbol positions corresponds to the mean value measured for at least 4 measurements in each condition, and the height of the errorbars to the associated  standard deviation.}
    The dotted line corresponds to $\Delta P_{\mathrm{pen}} = \Delta P_{L,\mathrm{max}}$.  }
    \label{fig:exp_th_carre_1}
\end{figure}

{To test this hypothesis, we carried out forced imbibition experiments for five mesh geometries and two fluids without yield stress: water, and a dilute Carbopol suspension ($c = 0.05$\% w/w $< c^*$).}
{In each case, we compare the experimentally measured penetration pressure $\Delta P_{\mathrm{pen}}$ against the maximal Laplace pressure $\Delta P_{L,\mathrm{max}}$, obtained from Eq.~\eqref{eq:Jamali} using $\Gamma=\SI{72}{\milli\newton\per\metre}$ for water, $\Gamma=\SI{63}{\milli\newton\per\metre}$ for Carbopol suspensions \cite{Jorgensen, boujlel2013}, $\theta=\ang{90}$, $m=m_{\mathrm{real}}$ and $\ell = e$ (Fig.~\ref{fig:exp_th_carre_1}).}
{The good overall agreement between the experimental and theoretical values confirms that, in the absence of yield-stress, the penetration of a fluid through a hydrophobic woven mesh is entirely determined by capillarity.}

\subsubsection{Penetration pressure for a yield-stress fluid}

{We now turn to the effect of yield-stress.}
{To this end, we performed quasi-static forced imbibition experiments for Carbopol solutions with yield-stress $\sigma_y$ ranging from $0$ to $\SI{85}{\pascal}$, through five mesh geometries, and systematically measured the penetration pressure $\Delta P_{\mathrm{pen}}$ (Fig.~\ref{fig:p_pen_sigma_y}).}
{The measurements at $\sigma_y = \SI{0}{\pascal}$ correspond to the ones presented above {for a dilute Carbopol suspension} (Fig.~\ref{fig:exp_th_carre_1}).}

\begin{figure}
 \centering
 \includegraphics[trim = 0.2cm 0.2cm 0.2cm 0.2cm, clip, width=\columnwidth]{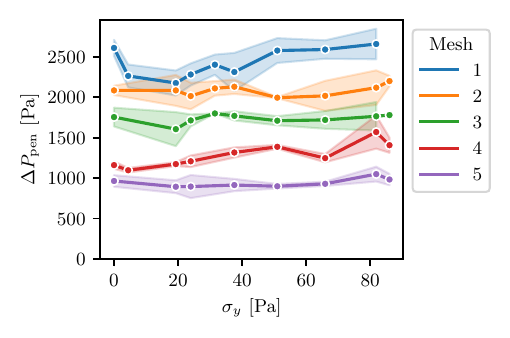}
 \caption{Penetration pressure $\Delta P_{\mathrm{pen}}$ as a function of yield stress $\sigma_y$, for the five meshes used.
 {Each point represents the average value for a set of parameters ($\sigma_y$, $m$, $e$) over 2 to 10 experiments. The lighter-colored areas indicating the corresponding standard deviations. }}
 \label{fig:p_pen_sigma_y}
\end{figure}

{We first note that the penetration pressure mostly depends on the mesh geometry.}
{The smaller the pore size, the larger the penetration pressure.}
{More precisely, the major difference between any two geometries seems to be largely explained by the behavior in the absence of yield stress, at $\sigma_y = \SI{0}{\pascal}$, where the penetration is governed by the Laplace pressure $\Delta P_{L, \mathrm{max}}$.}
{This observation suggests that we can separate the penetration pressure into a capillary and a yield-stress contribution,}
\begin{equation}
   \Delta P_{\mathrm{pen}} = \Delta P_{L, \mathrm{max}} + \Delta P_{\sigma_y}.
   \label{eq:yield_and_capillary}
\end{equation}

\noindent
{In such case, we estimate from a modified Darcy law the additional pressure required to make a yield-stress fluid flow through a pore of length $e$ and diameter $m_{\mathrm{real}}$\cite{Chevalier,Shahsavari},}
\begin{equation}
    \Delta P_{\sigma_y} = \beta \frac{e}{m_{\mathrm{real}}} \sigma_y,
    \label{eq:pression_seuil}
\end{equation}
{where $\beta$ is a numerical factor set by the geometry.}
{In Fig.~\ref{fig:p_seuil_sigma_y}, we plot the yield-stress contribution, $\Delta P_{\sigma_y}$, {obtained from Eq.~(\ref{eq:yield_and_capillary})},  as a function of the rescaled yield stress $(e/m_{\mathrm{real}})\sigma_y$.}

Although measurement uncertainties {dominate}, 
{the measured $\Delta P_{\sigma_y}$ is somehow consistent with a minimal flow model with a geometric prefactor $\beta =4$ (Eq.~\eqref{eq:pression_seuil}), comparable to other estimations in the literature \cite{Chevalier,Shahsavari}.}

\begin{figure}[t]
 \centering
 \includegraphics[trim = 0.2cm 0.2cm 0.2cm 0.2cm, clip, width=\columnwidth]{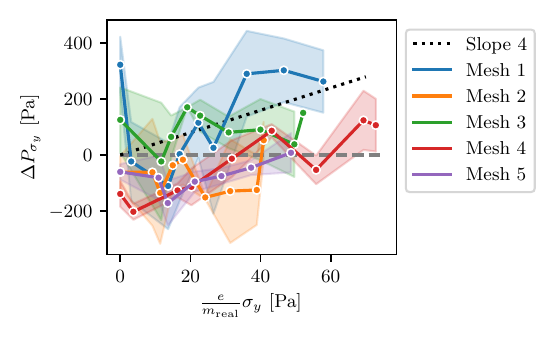}
 \caption{Yield-stress contribution to the penetration pressure $\Delta P_{\sigma_y}$ as a function of the rescaled yield stress $(e/m_{\mathrm{real}})\sigma_y$, for the five mesh geometries.
 The dotted line corresponds to $\beta = 4$ in Eq.~\eqref{eq:pression_seuil}
 {Each point represents the average value for a set of parameters ($\sigma_y$, $m$, $e$) over 2 to 10 experiments. The lighter-colored areas indicating the corresponding standard deviations.}}
 \label{fig:p_seuil_sigma_y}
\end{figure}

{We note, however, that the yield-stress contribution remains one order of magnitude lower than the capillary contribution $\Delta P_{L, \mathrm{max}}$.
The former scales with the yield stress, $\approx 10^2 \SI{}{\pascal}$ in our case, and the latter is of the order of $10^3 \SI{}{\pascal}$ (Fig.~\ref{fig:exp_th_carre_1}). 
For the range of yield stresses and mesh geometries investigated, the effect of yield-stress on the overall threshold penetration pressure is therefore quantifiable, but marginal.} {In order to observe a significant effect of the yield stress on the penetration pressure, i.e., $\Delta P_{L, \mathrm{max}} \sim \Delta P_{\sigma_y}$, a yield stress of at least $500-1000$ Pa would have been required for mesh sizes ranging from 60 to 200~$\mu$m.}

\subsection{Local penetration phenomenology is controlled by a priming pressure}

\subsubsection{A complex phenomenology}

{In contrast, the presence of a yield-stress plays a significant role in the imbibition phenomenology.
On the one hand, when the yield-stress is sufficiently high, the fluid inside adjacent pores coalesce when passing through the mesh (Fig~\ref{fig:grille} and Fig~\ref{fig:burstcoal}a).}
{On the other hand, in the absence of yield-stress, the fluid bursts through a single pore without touching the neighboring ones (Fig~\ref{fig:burstcoal}b).}
We identify these phenomena from changes in the light reflection pattern on the fluid-air interface.
When the fluid in adjacent pores merge, a lens effect makes the fibers below the fluid appear thicker (Fig~\ref{fig:burstcoal}a).
In turn, when the fluid goes through an isolated pore, the fluid–air interface becomes wider and blurry, indicating that the fluid has moved out of the focal plane (Fig~\ref{fig:burstcoal}b). 

{
To elucidate this transition, we investigated the penetration phenomenology of fluids with varying yield-stress values, through 5 different geometries, and gathered the results in a phase diagram (Fig.~\ref{fig:burstcoal}c).}
We observe that, in the absence of yield stress, the fluid systematically bursts through the mesh, regardless of the mesh geometry.
Conversely, beyond a yield stress of approximately \SI{50}{\pascal}, the fluid systematically coalesces during penetration. 
At intermediate yield stress values, the geometry of the mesh plays a significant role: burst dynamics dominate in smaller pores, and coalescence prevails in larger pores.
{The presence of a yield stress thus combines with the pore geometry to completely alter the penetration dynamics.}

\begin{figure}[h]
 \centering
 \includegraphics[width=\columnwidth]{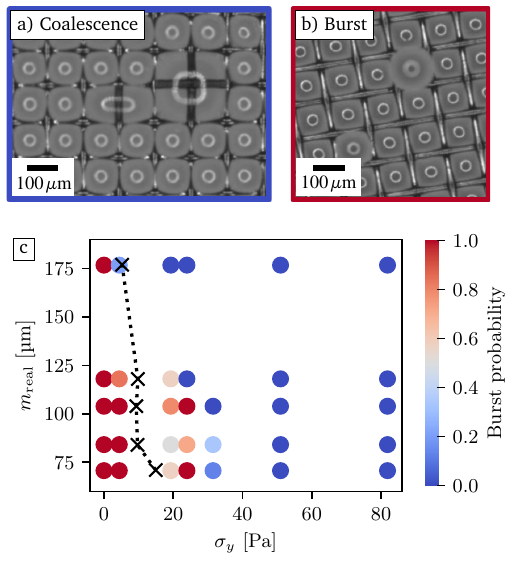}
 \caption{ 
 Snapshots of the penetration phenomenology with mesh 2.
 (a) The fluid coalesces with adjacent pores as it passes through. 
 (b) The fluid bursts through a single pore 
 (c) Phase diagram of the burst occurrences, as a function of the yield-stress $\sigma_y$ and the pore size $m_{\mathrm{real}}$.
 Each point represents the measured burst probability.
For each point, at least four videos were recorded, and up to eight videos in the transition zone where both burst and coalescence may occur.
The black crosses correspond to Eq.~\eqref{eq:critical_yield}, the dotted lines in between are a guide for the eyes.
 }
 \label{fig:burstcoal}
\end{figure}

\subsubsection{Fluid retraction controls penetration dynamics}

{In the previous section, we showed that the penetration pressure was largely controlled by the Laplace pressure from Eq.~\eqref{eq:Jamali}: once the fluid inside a pore reaches the immersion angle $\alpha_c$, at which the pressure equals the maximum Laplace pressure $\Delta P_{L,\mathrm{max}}$, it flows through the mesh. 
If the mesh was perfectly homogeneous, we would thus expect penetration to occur simultaneously in all the pores.}
The fact that the fluid instead goes through an isolated pore therefore points to the presence of mesh imperfections, such as variations in pore size or heterogeneities in surface wettability, which locally lowers the Laplace pressure
{
\cite{xiao2025quasi}}.

{When the maximal Laplace pressure is reached in this specific pore (Fig.~\ref{fig:jamali_2}, orange square), the fluid starts flowing through it {which} leads to a drop in Laplace pressure (Fig.~\ref{fig:jamali_2}, orange circle).
In the neighboring pores, however, the maximal Laplace pressure has not been reached (Fig.~\ref{fig:jamali_2}, blue square).
We then expect two possible behaviours.
Either the Laplace pressure inside the neighboring pores also drops, or it remains constant.
In the former case, the fluid retracts from the neighboring pores and empties into the unstable pore.
The fluid then flows through a unique pore, which corresponds to the burst scenario we observed at low yield-stress and small pore size (Fig.~\ref{fig:burstcoal}c).
In the latter case, the fluid keeps advancing in the unstable pore without retracting from the neighboring ones, until they coalesce.
}

\begin{figure}[ht]
    \centering
    \includegraphics[width=\columnwidth]{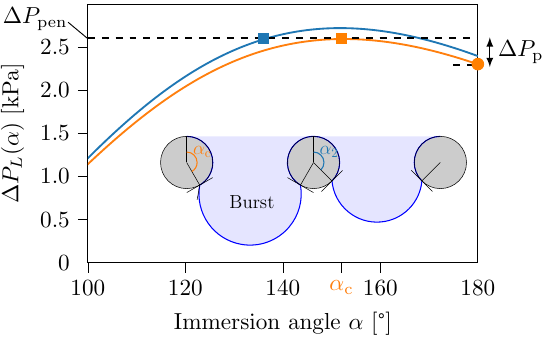}
%
%
    \caption{Main: Laplace pressure as a function of the immersion angle for two pores differing in size or wetting properties. One of them (orange curve) is characterized by a smallest capillary pressure. Inset: Illustration of {meniscus advancement} in two pores with different capillary pressures. The liquid goes through the pore with the smallest maximum pressure.}
    \label{fig:jamali_2}
\end{figure}

{A closer look at the light reflection pattern around the penetration time supports this simple model {of the two possible behaviours} (Fig.~\ref{retractation}).}
Step-by-step video analysis indeed shows how the reflection pattern reveals the interface position in each pore (Fig. \ref{fig:grille}).
The pattern is square-like at low pressure, when the immersion angle $\alpha$ is low (Fig.~\ref{fig:grille}d), and becomes circular as the interface advances through the pore (Fig.~\ref{fig:grille}e).
At low yield stress, as the fluid bursts through a single pore, the reflection pattern in neighboring pores rapidly turns back to a square-like shape, akin to the early stage of fluid advancement (Fig.~\ref{retractation} {top}).
This indicates that, as the fluid flows through a pore, it simultaneously retracts in the neighboring ones. {This phenomenology is reminiscent of Haines jumps that have been observed }{when Newtonian fluids penetrate hydrophobic porous media \cite{jamali2019penetration, haines1930studies, armstrong2013interfacial, moebius2012interfacial,sun2019haines}.}
Conversely, when the fluid of neighboring pores coalesces at high yield stress, the reflection pattern in the rest of the mesh remains circular (Fig.~\ref{retractation} {bottom}).
The progression of the fluid in each pore is therefore decoupled from that in neighboring pores. 
The pores do not interact and behave as if disconnected from each other.
For intermediate yield stress values, although no fluid retraction is observed during the first coalescence event, {retraction} gradually occurs as the coalescence includes more and more neighbors (Fig.~\ref{coalescences_suite_retractation}).

These observations are consistent with a drop of Laplace pressure, which \ML{pumps the neighbouring fluid. We thus name this pressure drop the \emph{priming pressure}}.
As the size of the coalescent drop increases, its curvature decreases and lowers the inner Laplace pressure.
The pressure difference, between the fluid in the coalescent drop and the neighboring pores, thus increases.
Eventually, it becomes large enough for the fluid in the neighboring pores to empty into the coalescent drop, in other words to retract, even in the presence of a yield stress. 
{Overall, this suggests that the penetration phenomenology is determined by a retraction process, which itself is governed by a competition between the yield stress and \ML{this} priming pressure.}

\begin{figure}[t]
    \centering
    \includegraphics[width=\columnwidth]{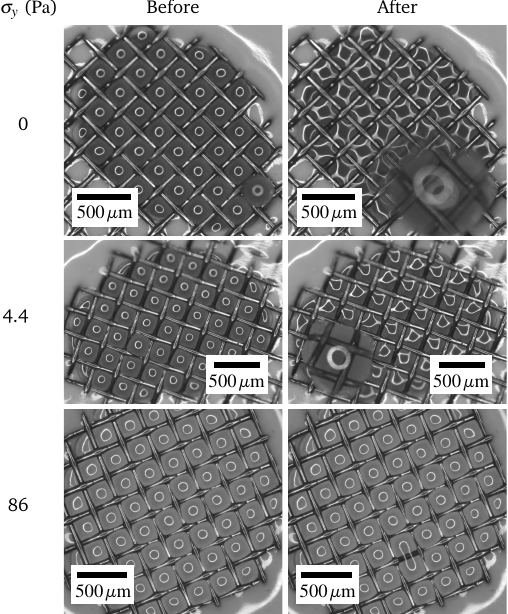}
    \caption{Snapshots from experimental videos obtained {with mesh 5} and {three different yield stresses (\SI{0}{\pascal}, \SI{4.4}{\pascal}, \SI{86}{\pascal})}. Snapshots referenced as `before' are taken just before the start of the flow and those referenced as `after' are taken just after. {The comparison of the reflection patterns between two snapshots (before/after) gives information about the existence or not of fluid retraction}.}
    \label{retractation}
\end{figure}

\begin{figure}[ht]
\includegraphics[width=\columnwidth]{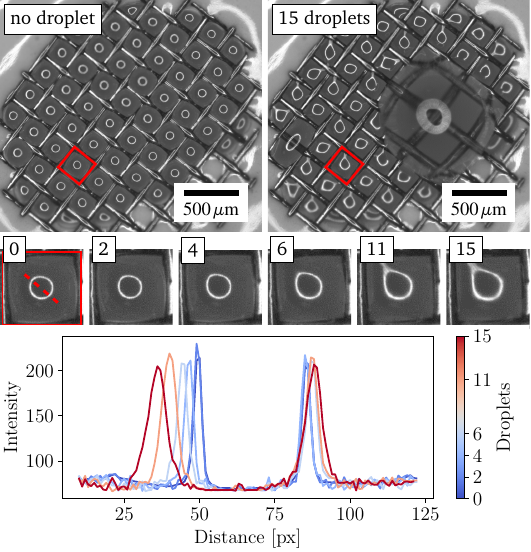}
    \caption{Snapshots from experimental videos with {mesh 5} and a fluid with a yield stress of \SI{24}{\pascal}, comparing fluid retraction before the first coalescence and when the coalescence cluster is formed by 15 droplets. The sequence of numbered images in the center of the figure displays the different stages of retraction in a selected pore (red square), as the coalescence cluster expands (for each image, we indicate the number of droplets in the coalescence cluster). The graph shows the evolution of light intensity along the red axis indicated in image 0, at each stage (the two peaks correspond to the edges of the white circle). }
    \label{coalescences_suite_retractation}
\end{figure}

\subsubsection{Priming pressure vs yield stress}

{To test this hypothesis, we further investigate the competition between yield stress and \ML{a} priming pressure, which we define as the difference between the Laplace pressure at penetration $\Delta P_{L,\mathrm{max}}$, and the Laplace pressure inside the unstable pore when the immersion angle reaches coalescence}
\begin{equation}
\Delta P_p=  \Delta P_{L,\mathrm{max}} -\Delta P_L(\alpha_{\mathrm{coalescence}}),
\end{equation}
{with $\alpha_{\mathrm{coalescence}}=180^\circ$  for a fluid-fiber contact angle $\theta=90^\circ$. Priming pressure $\Delta P_p$, defined as such, corresponds to the maximal pressure difference between the unstable pore and an adjacent one. 
If this pressure is insufficient to drive fluid retraction, imbibition occurs through a coalescence process.}

Assuming that the typical distance between the center of two adjacent pores is $m+d$, the resulting pressure gradient is $\nabla P_p=\Delta P_p/(m+d)$.
In the absence of yield stress, this pressure gradient drives the fluid retraction observed in the neighboring pores and is responsible for the burst regime.
In presence of a yield stress, however, a flow can only occur if this pressure gradient overcomes $\nabla P_y$, the pressure gradient required to advance the fluid through a pore of typical size $m$.
{In a fluid at rest}, it would be equal to $4\sigma_y/m$ for a cylindrical pore of diameter $m$ (see Supplementary Section 2 for details).
Here, as the fluid has already partially flowed through the pore, reverting its flowing direction implies $\nabla P_y=8\sigma_y/m$ \cite{Jorgensen}.

Following this model, fluid retraction systematically occurs when $\nabla P_p > \nabla P_y$.
In other words, for a given mesh geometry, the fluid only flows from the neighboring pores into the unstable one if its yield stress $\sigma_y$ is lower than the critical value
\begin{equation}
   \sigma_{yc} = \frac{m}{8(m+d)}\Delta P_p(m,\ell),
   \label{eq:critical_yield}
\end{equation}
which depends on the mesh geometry mainly through the priming pressure.
In Fig.~\ref{fig:burstcoal}c, we placed the point corresponding to $\sigma_y=\sigma_{yc}$ for each of the five mesh geometries we studied (black crosses).
We note that, as predicted by our model, this criteria clearly separates the region where the probability to burst is one from the rest of the phase diagram.

{This remarkable agreement demonstrates that, while the penetration pressure is set by capillary forces, the imbibition phenomenology is controlled by a tight interplay between the yield stress and a previously unreported priming pressure.}

\section*{Conclusions}

{In this work, we investigated the forced imbibition of a hydrophobic mesh by a water-based yield-stress fluid, for a range of mesh geometry and yield-stress values.
Combining pressure measurements and direct visualization techniques, we revealed the subtle way the yield-stress influences imbibition mechanics.
First, we focused on the pressure required to force the fluid through the mesh.
We measured the yield-stress contribution, and showed {that} it was minor compared to the capillary contribution.
Then, we turned to the imbibition pattern. 
We revealed that the manner the fluid goes through the mesh, either through a single pore or a collection of pores, is governed by an interplay between yield-stress and mesh geometry.
We explained it by detailing local pressure gradients, and unveiled the existence of a priming pressure, which we used to establish a predictive criterion for the penetration pattern as a function of yield-stress.}

{More generally, the penetration of a fluid into a non-wetting 2D mesh is controlled by two capillary pressures: the maximum Laplace pressure $\Delta P_{L, \mathrm{max}}$ and the priming pressure $\Delta P_p$. The first one sets the pressure at which the penetration occurs and the second one is responsible for the penetration pattern, i.e., through a single pore {with a mechanism similar to a Haines instability \cite{haines1930studies, armstrong2013interfacial, moebius2012interfacial,sun2019haines}}, or homogeneously through many pores. {In the case of the penetration of 2D mesh by} a yield-stress fluid, these two pressures have to be compared to the yield-stress value, giving two Bingham capillary numbers\cite{bertola2008,Martouzet2021}, {also known as plastocapillary numbers\cite{sanjay2021,jalaal2021}}, $B_{c,L}=\sigma_y/\Delta P_{L, \mathrm{max}}$ and $B_{c,p}=\sigma_y/\Delta P_p$. In the present study, $B_{c,L}\ll 1$ whereas $B_{c,p}\sim 1$, indicating that the yield stress does not affect the threshold penetration pressure but greatly influences on the penetration pattern.}
{In this regime, the yield stress is large enough to prevent Haines instability and thus ensure a more homogeneous penetration of the fluid in the porous matrix, while keeping the threshold penetration pressure almost as low as capillarity allows. It can be considered {as} the optimal regime for applications seeking homogeneous penetration at minimal pressure. In this case, using a low yield stress fluid is beneficial with respect to a Newtonian fluid.}

{An attractive perspective of this work would be to study the penetration of a yield-stress fluid into {non-wetting model porous media on  microfluidic chips \cite{armstrong2013interfacial, moebius2012interfacial} or} into a few 2D meshes associated in series \cite{jamali2019penetration}. {The latter} would be a first step to model the penetration of 3D fibrous porous medium by a non-wetting complex fluid. Another perspective could be to go towards industrial applications and their more specific complex fluids such as cement paste, paints and cosmetics.  For instance, incorporating grains inside the yield-stress fluid would enable the exploration of the coupling between rheology, mesh geometry, grain size and solid fraction \cite{Roussel_clog}.}

\section*{Author contributions}
{MB designed the experimental set-up and conducted all experiments under the supervision of CB, ML, LV, and NB. MB analyzed the data, and CB developed the model to interpret it. All authors contributed equally to writing the article. Funding acquisition by CB, ML and LV.} 

\section*{Conflicts of interest}

There are no conflicts to declare.

\section*{Data availability}
Data are available at \url{https://doi.org/10.5281/zenodo.16412539} and include:
\begin{itemize}
\item \verb|Donnees_manip_compression_QS_sup.csv| containing the experimental conditions and data for the pressure measurements;
\item \verb|Figures_Article.ipynb|, the python code that generates the figures from the experimental data;
\item the movies showing the penetration of 2D hydrophobic meshes by water-based fluids (\verb|fig2.avi|, \verb|fig7a.avi|, \verb|fig7b.avi|, \verb|fig9_0Pa.avi|, \verb|fig9_4p4Pa.avi|, \verb|fig9_86Pa.avi|, \verb|fig10|) from which images shown in the Figures have been extracted;  
\item intensity profiles (\verb|goutte_0.csv|, \verb|goutte_1.csv| to \verb|goutte_6.csv|) used to plot the graph in Figure 10;
\item figures (pdf files).
 \end{itemize}


\section*{Acknowledgements}
{The authors acknowledge Saint-Gobain Research for  financial support, in particular the CIFRE grant of Manon Bourgade.  The authors also thank Gilles Simon and Matthieu Mercury for technical support, and Solenn Moro, Julie Godefroid and Paul Jacquet for fruitful discussions. The authors thank Changwoo Bae and Anne-Laure Biance for their set up and assistance in measuring the microscopic contact angles.}




\renewcommand\refname{References}

\bibliography{rsc} 
\bibliographystyle{rsc} 

\newpage
\section*{Supplementary information}

\subsection*{{1. Hydrophobic 2d meshes}} \label{meshes_supmat}

To attach the mesh to the syringe {(3cc syringes from \textsc{Nordson})}, the mesh is first stretched, and an adhesive (\textsc{Loctite SI 5398}) is carefully applied to the syringe tip. 
The mesh is then gently pressed onto the adhesive and left to dry for about 15 minutes.
Each assembly only allows for a single measurement because the hydrophobic coating on the meshes is delicate and degrades after use. 
Therefore, after each measurement, the used mesh is removed from the syringe and replaced with a new one for subsequent tests. 

\begin{table}[h]
    \centering
    \begin{tabular}{@{}cc@{}}
                \toprule
			Mesh \# & Reference \\\midrule
                1& SEFAR NITEX 03-64/45 hydrophobic 150 cm \\
                2& SEFAR ACOUSTIC 6-85 BSY \\
                3& SEFAR ACOUSTIC 6-105 BSY \\
                4& SEFAR NITEX 03-125/45 black hydrophobic 102 cm\\
                5& SEFAR NITEX 03-190/57 black hydrophobic 102 cm \\
                \bottomrule 
            \end{tabular}
    \caption{Hydrophobic mesh references.}
    \label{tab:sefar_mesh_ref}
\end{table}

Among the five SEFAR meshes used in this study, four (mesh 2 to mesh 5) have a similar surface treatment, while the mesh 1 seems slightly different. {To the eye}, mesh 1 is white, while the others are black, indicating a variation in surface treatment. 
{Furthermore}, a comparison of mesh flexibility (see Fig. \ref{fig:mesh_comp}) shows that {mesh 1} is more flexible than the others, making it more susceptible to deformation.
Such deformations could explain some of the data variability. 

\begin{figure}[h!]
    \centering
    \includegraphics[width=\columnwidth]{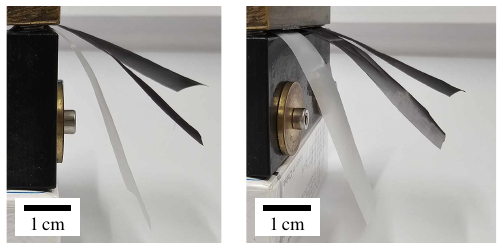}
%
%
    \caption{Mesh flexibility comparison using three mesh samples of the same size and similar radius fibers. From left to right: {mesh 1 (white), mesh 2, and mesh 3. Mesh 1} is significantly less rigid than the other two.}
    \label{fig:mesh_comp}
\end{figure}

\subsection*{2. Estimation of $P_{\mathrm{mesh}}$ from the applied pressure $P_0$}


\subsubsection*{a) Hydrostatic pressure}

The syringe orientation varied between experiments: upward orientation is used with yield-stress fluids and downward without yield stress \ML{(See Fig.\ref{fig:schema_compression_visualisation})}. In this latter case, the existence of air bubbles with the upward orientation disturbs the measurements so that the syringe has to be reversed.  
{In order to link the air compressor $P_0$ and the pressure at the mesh interface $P_{\mathrm{mesh}}$, we have to take into account the hydrostatic pressure $\Delta P_h = \rho g h$, where $\rho$ is the fluid density, $g$ the gravitational acceleration, and $h$ the difference between the altitude of the surface of the fluid where $P_0$ is applied and the altitude of the mesh.
When the syringe is oriented downward as in Fig 1. of the main text, $h>0$ thus the hydrostatic contribution is positive. Conversely, when the syringe is oriented upward, $h<0$, and the hydrostatic contribution to $P_{\mathrm{mesh}}$ is negative. Once the hydrostatic contribution taken into account, we could not find significant differences between experiments performed upward or downward, other parameters being equal.
}

\begin{figure}[b]
    \centering
    \includegraphics[width=\columnwidth]{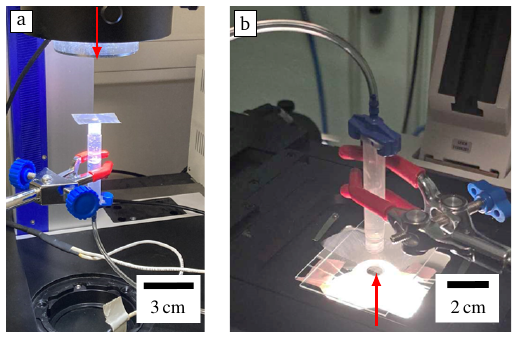}
%
    \caption{\ML{Pictures of the two experimental configurations : a) upward orientation and b) downward orientation. The red arrow shows the light direction and the camera location.} }
    \label{fig:schema_compression_visualisation}
\end{figure}

\begin{figure}[h]
    \centering
    \includegraphics[width=\columnwidth]{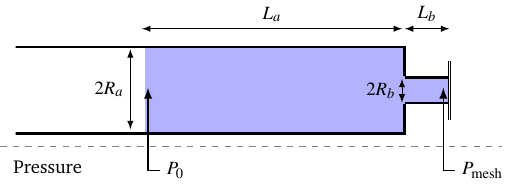}
%
%
%
%
%
%
    \caption{Simplified syringe used for the experiments and definition of useful dimensions. {$L_a$ depends on} the amount of fluid inside the syringe and was measured for each experiment (around $\SI{3}{\centi\metre}$). }
    \label{fig:schema_seringue_ecoulement}
\end{figure}
\subsection*{b) Extra pressure due to yield stress}
In the case of a yield-stress fluid, an extra pressure difference $\Delta P_y$ has to be applied to drive the fluid along the syringe such as: {$P_{\mathrm{mesh}} = P_0 + \Delta P_h -\Delta P_y$.} 

To find the expression of $\Delta P_y$, {we solve Navier-Stokes equation assuming incompressible  and quasi-static flow}:
\begin{equation}
\rho \vec{v}.\vec\nabla\vec{v}=\vec\nabla\sigma -\vec\nabla P \quad \text{and} \quad \vec{\nabla}.\vec{v}=0,
\end{equation}
where $\sigma$ is the stress tensor.
In the case of a cylinder of length $L$ and radius $R$ such as $L\gg R$, the invariance and symmetry of the flow coupled to its incompressibility lead to:
\begin{equation}
\vec{v}=v_z(r)\vec{e_z} \quad \text{and} \quad \frac{1}{r}\frac{\partial (r\sigma_{z,r})}{\partial r}-\frac{\partial P}{\partial z}=0
\end{equation} 
in cylindrical coordinates. Here, the pressure only depends on the $z$ coordinate and the stress tensor on $r$. The integration of Navier-Stokes equation on $r$ leads to the following relation between the pressure gradient and the stress tensor: $\sigma_{z,r}(r)=\frac{r}{2}\frac{\partial P}{\partial z}$. The stress is zero at the center of the cylinder and maximum at the edge of the cylinder. In a quasi-static condition, the minimum pressure gradient to apply in order to drive a yield-stress fluid along the cylinder is then: $\frac{\partial P}{\partial z}=\frac{2\sigma_y}{R}$. Consequently, the minimum pressure difference to drive a yield stress fluid along a cylinder of length $L$ is:  $\Delta P_y=\frac{2L\sigma_y}{R}$.
In our experimental set-up, the syringe is composed of two consecutive cylinders (see Fig. \ref{fig:schema_seringue_ecoulement}) of {respective lengths ($L_a=\SI{3}{\centi\metre}$, $L_b=\SI{1}{\centi\metre}$) and radii ($R_a=\SI{4.75}{\milli\metre}$, $R_b=\SI{1}{\milli\metre}$)}, so the minimal difference pressure $\Delta P_s$ is given by:
\begin{equation}
{\Delta P_y =2\sigma_y \left(\frac{L_a}{R_a}+\frac{L_b}{R_b}\right)}
\label{ecoulement_th}
\end{equation}
In Fig. \ref{fig:pression_ecoulement_graphe}, we plot the measured pressure difference $\Delta P_y$ as a function of the yield stress. {Our data agrees well with a linear dependence between the two quantities}. And the slope (here 33.6) is in a good agreement with the predicted one (32.6).

\begin{figure}
 \centering
 \includegraphics[trim = 0.2cm 0.2cm 0.2cm 0.2cm, clip, width=\columnwidth]{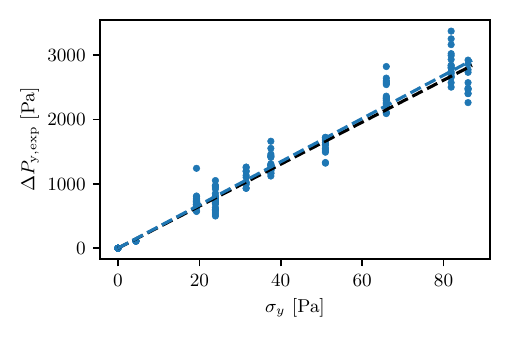}
 \caption{Pressure difference $\Delta P_{\mathrm{y,exp}}$ needed to drive the yield stress fluid to the end of the syringe as a function of the yield-stress $\sigma_y$. The black dashed line corresponds to the theoretical pressure difference from Eq.~\ref{ecoulement_th}. }
 \label{fig:pression_ecoulement_graphe}
\end{figure}





\subsection*{3. Detailed experimental steps}
\ML{Here are the different steps of an experiment with our quasi-static setup : 
\begin{itemize}
    \item we glue a piece of the chosen mesh to the end of the syringe,
    \item we partially fill the syringe by hand with the model yield-stress fluid, being careful not to put too much bubbles inside the yield-stress fluid, especially with the highest yield stresses used in this work,
    \item we connect the air compressor to the other end of the syringe and place the syringe in front of the camera, upward or downward depending on the yield stress of the fluid,
    \item we start the video recording of the experiment and then manually increase the applied pressure until the fluid starts flowing through the mesh, at a slow rate of about \SI{10}{\pascal \per\second} to keep a quasi-static flow,
    \item the experiment is stopped and the exact fluid height $L_a$ is measured for post-experiment analysis. 
\end{itemize}}

\subsection*{4. Carbopol preparation protocol}
Carbopol suspensions are prepared as follows:
\begin{itemize}
\item A specified amount of polymer powder, corresponding to a mass concentration between 0.08\% and 3\%, is added to \SI{40}{\gram} of pure water heated to approximately \SI{50}{\celsius} {in order to accelerate dissolution}.

\item The solution is stirred for 30 minutes until the powder is fully dissolved. To prevent evaporation, the container is sealed with parafilm during this phase.

\item After stirring, the solution has an acidic pH of around 3. Sodium hydroxide is added to neutralize the pH to 7 $\pm$ 1 (monitored using pH paper). For instance, for a 1\% mass concentration Carbopol suspension (\SI{0.4}{\gram} Ultrez 10 powder), approximately \SI{275}{\micro\litre} of 10~M NaOH is required. This step causes the polymer network to swell{. At sufficient mass concentration, $c>c^{*}\approx 0.09\%$ w/w, the microgels jam, endowing a} yield stress to the suspension.

\item Once neutralized, the suspension is strongly stirred using a mixer (RW20, Ika) at 900 rpm for 4 hours to homogenize the microstructure.

\item Finally, the suspension is centrifuged to remove air bubbles introduced during stirring, particularly in the second phase.

\end{itemize}

The suspension is stored in a refrigerator.

\subsection*{5. Flow curve and $\sigma_y$ measurement}

To characterise the yield stress of Carbopol suspensions, flow curves are measured using a stress-controlled rheometer (Anton Paar Physica MCR 302) with a parallel-plate geometry.
To prevent Carbopol from slipping, a layer of P320 sandpaper added to the surfaces of both plates is used. The gap is set to \SI{1}{\milli\metre}. 
The following protocol is used:
\begin{itemize}
\item Pre-shear at $\dot\gamma = \SI{100}{\s^{-1}}$ for \SI{1}{\minute} to ensure reproducible starting state.
\item Decreasing shear rate ramp from $\dot\gamma = \SI{100}{\s^{-1}}$ to $\SI{0.01}{\s^{-1}}$, measuring 10 points per decade, {waiting \SI{30}{\second}} at each point to ensure steady state.
\item Increasing shear rate ramp from $\dot\gamma = \SI{0.01}{\s^{-1}}$ to $\SI{100}{\s^{-1}}$, measuring 10 points per decade, {waiting \SI{30}{\second}} at each point to ensure steady state.
\end{itemize}

\begin{figure}
\tikzsetnextfilename{fit_hb}
\includegraphics[width=\columnwidth]{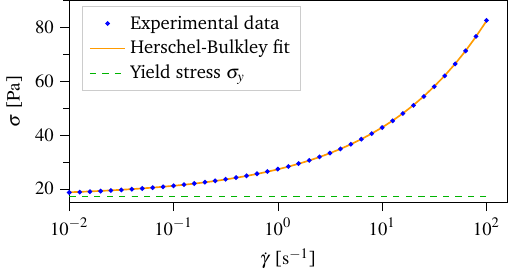}

    \caption{Flow curve {with decreasing shear rate ramp} of a 0.2\% w/w Carbopol U10 suspension (blue dots), fitted with the Herschel-Bulkley model (orange curve), and yield stress from the fit (green dashed line).}
    \label{fig:fit_hb}
\end{figure}

The flow curves are then fitted using the Herschel-Bulkley (HB) model: 
\begin{equation}
  \begin{cases}
    \sigma = \sigma_y + K\dot\gamma^n & \text{if } \sigma \geq \sigma_y \\
    \dot\gamma = 0  & \text{if } \sigma < \sigma_y \\
  \end{cases}
\end{equation}
where $\sigma_y$ is the yield stress, $K$ the consistency and $n$ the exponent of the HB law.
The fitting is performed using the \texttt{curve\_fit} function from the \texttt{scipy} library \cite{Scipy}. An example of flow curve and HB fit is shown in Fig. \ref{fig:fit_hb}. As Carbopol is a model yield-stress fluid, the fit closely matches the experimental data with an error of approximately 0.1\%.
The primary source of uncertainty comes from the measurement itself.
Indeed, since the gel is manually deposited onto the lower plate using a spatula, slight variations in volume can occur between experiments. 
After positioning the upper plate \SI{1}{\milli\metre} above the lower plate, any excess Carbopol overflowing the edge is carefully removed with absorbent paper. 
However, this step introduces slight variations between measurements, particularly in the residual edge thickness, which may affect the stress measured by the rheometer.
As a result, repeated flow curve measurements on the same gel can exhibit variations of about 10\%. For concentrations of Carbopol Ultrez 10 powder ranging from 0.1\% w/w to 1.2\% w/w,  we obtain $\sigma_y$ varying between 4.4$\pm 0.4$ and $86\pm 9$~Pa.

\subsection*{6. Contact angle measurement}

To quantify the hydrophobicity of the meshes, the contact angle formed by a water drop on the mesh surface was measured. 
Images, such as shown in Fig.~\ref{fig:contact_angle_meas} Left, were captured using a horizontally positioned camera, with a light source placed on the opposite side of the sample to enhance contrast. 
Image analysis was performed with ImageJ using the \href{https://bigwww.epfl.ch/demo/dropanalysis/}{\texttt{DropSnake}} module \cite{dropsnake}, enabling the contact angle measurement thanks to a precise identification of the interface between the mesh, the drop, and the air. This contact angle, denoted $\theta_{\textrm{macro}}$ corresponds to a macroscopic Cassie contact angle on a rough surface and its average value (on 4 drops) is: $\theta_{\textrm{macro}}=130^{\circ}\pm \ang{10}$. \ML{Moreover, as we inflate the drop during its deposition on the meshes, the macroscopic contact angle corresponds to an advancing angle. This measurement is consistent with the experience of a fluid penetrating a medium, since only advancing angles are involved.}

We also measure the \ML{advancing} contact angle on individual fibers. This angle, denoted $\theta_{\textrm{micro}}$, characterises the affinity of the liquid with the fiber in presence of the air, independently of the geometrical properties of mesh. Measuring contact angles on single fibers presents challenges due to their small diameters (tens of microns), requiring water droplets smaller than the fiber diameter. 
For instance, a droplet on a \SI{62}{\micro\metre} SEFAR fiber must have a maximum volume of approximately \SI{0.1}{\nano\litre}.
To minimize evaporation during imaging, the fiber was placed in a sealed, high-humidity chamber (near 100\%). 
Saltwater droplets were used, applied via a fine-tipped silanized glass pipette to ensure precise control of droplet size and placement. 
The setup allowed the deposition of microscopic droplets while preventing evaporation. 
A side-mounted objective captured images for subsequent contact angle measurement (see  Fig.~\ref{fig:contact_angle_meas} Right). The average value (on 5 drops) of the contact angle on single fiber is: $\theta_{\textrm{micro}}=95^{\circ}\pm \ang{10}$. \ML{This value is used to estimate the contact angle $\theta$ that appears in Eq.~\ref{eq:Jamali}. This is indeed only a rough estimation since $\theta$ is the contact angle on a flat surface and $\theta_{\textrm{micro}}$ is measured on a curved surface \cite{bormashenko2009,guilizzoni2011}. However, to minimize the effect of the surface curvature on the contact angle measurement, we consider droplets as small as possible. } 

\begin{figure}[b]
    \centering
    \includegraphics[width=\columnwidth]{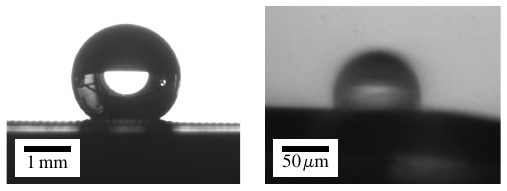}
%
%
%
    \caption{Macroscopic and microscopic contact angle measurement. Left: water drop on {mesh 2}. Right: water droplet on a single fiber of {mesh 4.}}
    \label{fig:contact_angle_meas}
\end{figure}

\end{document}